\newif\iffull
\fulltrue   

\iffull
	\documentclass[11pt]{article}
	\usepackage{fullpage}
	\def\squarebox#1{\hbox to #1{\hfill\vbox to #1{\vfill}}}
	\newcommand{\qed}{\hspace*{\fill}
			\vbox{\hrule\hbox{\vrule\squarebox{.667em}\vrule}\hrule}\smallskip}

	\newenvironment{proof}{\noindent{\bf Proof:~~}}{\(\qed\)}
\else
	\documentclass[acmsmall]{ec21acm}
	\setcopyright{rightsretained}
\fi

\usepackage{mathrsfs}
\usepackage{amsmath}
\usepackage{amsfonts}
\usepackage{paralist}
\usepackage{float}
\usepackage{color}
\usepackage{dsfont} 
\usepackage[most]{tcolorbox} 

\newtheorem{theorem}{Theorem}[section]

\newtheorem{lemma}[theorem]{Lemma}

\newtheorem{claim}[theorem]{Claim}

\usepackage{graphicx}
\newcommand{\eps}{\varepsilon}

\newcommand{\expct}{\mathbb{E}}
\newcommand{\one}{\mathds{1}}

\newcommand{\D}{\mathcal{D}}
\newcommand{\R}{\mathcal{R}}
\newcommand{\SP}{\mathcal{SP}}
\newcommand{\opt}{\textsf{OPT}}

\newcommand{\tD}{\widetilde{\D}}
\newcommand{\hatp}{\hat{p}}
\newcommand{\hatt}{\hat{t}}
\newcommand{\algdist}{\mathcal{C}}

\newcommand{\Rhigh}{\R^{\textsc{high}}}

\newcommand{\Rlow}{\R^{\textsc{low}}}

\newcommand{\SPhigh}{\SP^{\textsc{high}}}
\newcommand{\SPmid}{\SP^{\textsc{mid}}}
\newcommand{\SPlow}{\SP^{\textsc{low}}}

\newcommand{\maxone}{\textrm{max}}
\newcommand{\smax}{\textrm{max}_{(2)}}

\newcommand{\Xonemax}{X^1_{\textrm{max}}}
\newcommand{\Xtwomax}{X^2_{\textrm{max}}}
\newcommand{\Yonemax}{Y^1_{\textrm{max}}}
\newcommand{\Ytwomax}{Y^2_{\textrm{max}}}

\newcommand{\econ}{\mathbf{E}}
\newcommand{\Eopt}{\econ_{\opt}}
\newcommand{\Eeps}{\econ_{\eps}}

\newcommand{\econtilde}{\widetilde{\econ}}

\newcommand{\DG}[1][]{\D^{G_{#1}}}

\newcommand{\rev}[1][\econ]{\R_{#1}}

\newcommand{\sprice}[1][\econ]{\SP_{#1}}

\newcommand{\highrev}[1][\econ]{\Rhigh_{#1}}

\newcommand{\lowrev}[1][\econ]{\Rlow_{#1}}

\newcommand{\highsprice}[1][\econ]{\SPhigh_{#1}}
\newcommand{\midsprice}[1][\econ]{\SPmid_{#1}}
\newcommand{\lowsprice}[1][\econ]{\SPlow_{#1}}

\newcommand{\RR}{{\mathbb{R}}}

\usepackage{soul}

\title{Simple Economies are Almost Optimal}

\iffull
	\author{%
		Amir Ban%
		\thanks{Weizmann Institute of Science;
		\texttt{amir.ban@weizmann.ac.il}}
	\and
		Avi Cohen%
		\thanks{Tel Aviv University;
		\texttt{avicohen2@mail.tau.ac.il}}
	\and
		Shahar Dobzinski%
		\thanks{Weizmann Institute of Science;
		\texttt{shahar.dobzinski@weizmann.ac.il}}
	\and
		Itai Ashlagi%
		\thanks{Stanford University;
		\texttt{iashlagi@stanford.edu}}
	}
\else
	\author{Amir Ban}
	\email{amir.ban@weizmann.ac.il}
	\affiliation{%
		\institution{Weizmann Institute of Science}
		\city{Rehovot}
		\country{Israel}
	}

	\author{Avi Cohen}
	\email{avicohen2@mail.tau.ac.il}
	\affiliation{%
		\institution{Tel Aviv University}
		\city{Tel Aviv}
		\country{Israel}
	}

	\author{Shahar Dobzinski}
	\email{shahar.dobzinski@weizmann.ac.il}
	\affiliation{%
		\institution{Weizmann Institute of Science}
		\city{Rehovot}
		\country{Israel}
	}

	\author{Itai Ashlagi}
	\email{iashlagi@stanford.edu}
	\affiliation{%
		\institution{Stanford University}
		\city{Stanford}
		\country{USA}
	}
\fi

\iffull
\else

\fi

\begin{document}

\iffull
	\maketitle
	\begin{abstract}
Consider a seller that intends to auction some item. The seller can invest money and effort in advertising in different market segments in order to recruit $n$ bidders to the auction. Alternatively, the seller can have a much cheaper and focused marketing operation and recruit the same number of bidders from a single market segment. Which marketing operation should the seller choose?

More formally, let $D=\{\mathcal D_1,\ldots, \mathcal D_n\}$ be a set of 
distributions. Our main result shows that there is always $\mathcal D_i\in D$ such that the revenue that can be extracted from $n$ bidders, where the value of each is independently drawn from $\mathcal D_i$, is at least $\frac 1 2 \cdot (1-\frac 1 e)$ of the revenue that can be obtained by any possible mix of bidders, where the value of each bidder is drawn from some (possibly different) distribution that belongs to $D$.


We next consider situations in which the auctioneer cannot use the optimal auction and is required to use a second price auction. We show that there is always $\mathcal D_i\in D$ such that if the value of all bidders is independently drawn from $\mathcal D_i$ then running a second price auction guarantees a constant fraction of the revenue that can be obtained by a second-price auction by any possible mix of bidders. Finally, we show that for any $\varepsilon>0$ there exists a function $f$ that depends only on $\varepsilon$ (in particular, the function does not depend on $n$ or on the set $D$), such that recruiting $n$ bidders which have at most $f(\varepsilon)$ different distributions, all from $D$, guarantees $(1-\varepsilon)$-fraction of the revenue that can be obtained by a second-price auction by any possible mix of bidders.

\end{abstract}
\else
	\fancyhead{}
	\fancyfoot{}

	
	\copyrightyear{2021}
	\acmYear{2021}
	\acmConference[EC '21]{Proceedings of the 22nd ACM Conference on Economics and Computation}{July 18--23, 2021}{Budapest, Hungary}
	\acmBooktitle{Proceedings of the 22nd ACM Conference on Economics and Computation (EC '21), July 18--23, 2021, Budapest, Hungary}\acmDOI{10.1145/3465456.3467563}
	\acmISBN{978-1-4503-8554-1/21/07}
	

	

	\settopmatter{printacmref=true}
	\maketitle
\fi

\iffull
\section{Introduction}
\fi


Optimal mechanisms are often unnatural or hard to implement. Thus, an influential line of work in Algorithmic Mechanism Design attempts to develop suboptimal but simple mechanisms that are almost as good as optimal ones. Some of the numerous examples include \cite{HR09, CHMB10, R01, KW12}.


Other papers suggest to tackle the inapplicability of optimal auctions by approaching the problem from a different perspective: instead of changing the mechanism, change the market. The seminal paper of Bulow and Klemperer \cite{BK94} initiated this line of research: it shows that the optimal revenue that can be extracted from an economy with $n$ bidders whose values are drawn i.i.d. from some regular distribution $\mathcal D$ is at most the revenue of a second price auction with $n+1$ bidders, all drawn i.i.d. from $\mathcal D$. That is, recruiting a single additional bidder allows the auctioneer to use the simple second-price auction without losing much revenue. Follow-up papers presented approximate versions of similar statements in other settings, e.g., for certain non-regular distributions \cite{SS13}, when the distributions are not identical \cite{FLR19, HR09}, and when there are multiple heterogeneous items \cite{EFFTW17, FFR18}.

In this paper we suggest to explore a new market-changing approach. Consider a hypothetical scenario of a seller who intends to auction some item. The seller can invest money and effort in advertising in different market segments in order to recruit bidders to the auction. Alternatively, the seller can have a much cheaper and focused marketing operation and recruit the same number of bidders from a single market segment. Which marketing operation should the seller choose? Our goal is to compare the effectiveness of the different strategies.

More formally, let $D=\{\mathcal D_1,\ldots, \mathcal D_n\}$ be a set of 
distributions. An economy $\mathbf E$ consists of $n$ bidders, where the value of each bidder $i$ is independently drawn from some $\mathcal D\in D$. For each economy $\mathbf{E}$, let $\mathcal R_{\mathbf {E}}$ be the revenue of the optimal auction for the economy $\mathbf E$. The ideal revenue is defined to be $\max_{\mathbf E}\mathcal R_{\mathbf{E}}$.

Our goal is to determine whether there exists some distribution $\mathcal D\in D$ such that the revenue of the optimal auction for an economy where the values of the bidders are drawn i.i.d. from $\mathcal D$ (a \emph{homogeneous} economy) provides a good approximation to the ideal revenue of the economy. In other words, we would like to determine whether the revenue that can be extracted by recruiting all bidders from the same population is comparable to the revenue that can be generated by handpicking the bidders in a way that maximizes the revenue. Our main result shows that it is always possible to extract a constant fraction of the ideal revenue with a homogeneous economy:

\vspace{0.1in} \noindent \textbf{Theorem: }Let $D=\{\mathcal D_1,\ldots, \mathcal D_n\}$ be a set of distributions. There is a homogeneous economy $\mathbf E$ where the value of each bidder $i$ is independently drawn from the same distribution $\mathcal D\in D$, such that the revenue of an optimal auction for $\mathbf E$ is at least a $\left (\frac 1 2 \cdot (1-\frac 1 e)\right )$-fraction of the ideal revenue.

\vspace{0.1in} \noindent
\iffull
Note that we obtain this bound by running a simple second price auction with a reserve price in the homogeneous economy. It is not hard to see that sometimes no homogeneous mechanism can extract more than half of the ideal revenue\footnote{A naive and wrong solution would be to take the distribution $\mathcal D\in D$ that has the highest value in the support and have infinitely many bidders with values drawn from $\mathcal D$. However, this requires much more than $n$ bidders and thus is impractical and in particular infeasible in our case.}. Consider a distribution $\mathcal D_1$ that returns the value $1$ with probability $1$, and a distribution $\mathcal D_2$ that returns 0 with probability $1-\eps$ and $\frac 1 {n\cdot \eps}$ with probability $\eps$, for some very small $\eps>0$. The ideal revenue is (very close to) $2$, and it can be obtained by the economy in which the value of one bidder is drawn from the distribution $\mathcal D_1$ and the value of each one of the remaining $n-1$ bidders is to be drawn from $\mathcal D_2$. The optimal auction for this economy is a second price auction with a reserve price of $1$ for the bidder whose value is drawn from $\mathcal D_1$, and a reserve price of $\frac 1 {n\cdot \eps}$ for the remaining bidders. Finally, note that the revenue is at most $1$ if we choose the values of all $n$ bidders to be drawn i.i.d. from $\mathcal D_1$ or the values of all $n$ bidders to be drawn i.i.d. from $\mathcal D_2$. In both cases the social welfare is at most $1$, which gives us an immediate bound on the attainable revenue.

This establishes that homogeneous economies can always extract a constant fraction of the ideal revenue, and that sometimes they cannot do better\footnote{Interestingly, in Appendix~\ref{app-regular} we show that sometimes no homogeneous mechanism can extract more than half of the ideal revenue even if all distributions are regular.}.

\fi
The mechanism designer is often unable to freely choose an auction format. Thus, we now consider situations in which the mechanism designer is constrained to use a second price auction in all economies, either because, e.g., it is not allowed to tailor a mechanism based on the specifics of the distributions or because of the simplicity of a second price auction. For each economy $\mathbf{E}$. We prove that homogeneous markets approximate well the ideal second-price revenue.

\vspace{0.1in} \noindent \textbf{Theorem: }Let $D=\{\mathcal D_1,\ldots, \mathcal D_n\}$ be a set of distributions. There exists a homogeneous economy $\mathbf E$ where the value of each bidder is independently drawn from the same distribution $\mathcal D\in D$, such that the revenue that can be generated by a second price auction in $\mathbf E$ is at least a $c$-fraction of the ideal second-price revenue, for some constant $c>0$.

\vspace{0.1in} \noindent Next, we consider whether being able to recruit bidders from a small number of market segments (in contrast to just one, as in homogeneous markets) allows to significantly extract more revenue. We answer this question in the affirmative, at least when restricted to the second-price auction:

\vspace{0.1in} \noindent \textbf{Theorem: }Let $D=\{\mathcal D_1,\ldots, \mathcal D_n\}$ be a set of distributions. There exists some function $f$ that depends only on $\eps$, such that for every constant $\eps>0$ the following holds: there exists a set $D'\subseteq D$, $|D'|=f(\eps)$, and an economy $\mathbf E$ where the value of each bidder is drawn indpendently from some distribution $\mathcal D\in D'$, such that the revenue that can be generated by a second-price auction for $\mathbf E$ is at least a $\left (1-\eps \right )$-fraction of the ideal second-price revenue.

\vspace{0.1in} \noindent 
\iffull
To put it differently, for every constant $\eps>0$, there exists a simple economy that uses only a constant number of distributions (irrespective of the number of bidders $n$ and the set $D$) and extracts almost all the ideal second-price revenue\footnote{In fact, $f(\eps)=exp(\frac 1 \eps)$. We do not know whether this dependency is tight or maybe we can have $f(\eps)=poly(\frac 1 \eps)$. }. 

Approximating the ideal revenue of a second price auction with homogeneous economies is somewhat reminiscent of the team formation problem \cite{KR18}. In this problem we are given $n$ different candidates, each candidate $i$ is modeled as a random variable from some distribution $\mathcal D_i$ and the goal is to choose $n$ team members that will maximize the expected sum of values of the top $h$ chosen candidates. One important difference between our paper and \cite{KR18} is that their goal is to maximize the sum of the highest $h$ variables, which does not seem to easily imply -- if at all -- a bound on the expected second-highest value. Also related is a recent paper by Mehta et al. \cite{MNPR20}. In their case they consider a population of $n$ bidders and their goal is to select $k<n$ bidders that maximize the revenue of the second-price auction. The main focus of this paper is the computational complexity of this problem, and they show that under the planted clique hypothesis there is no constant factor algorithm that runs in polynomial time. They also provide a PTAS for maximizing the expectation of the highest value. Other papers (e.g., \cite{EL02}) consider the computational aspects of choosing $k$ out of $n$ random variables to minimize the expected minimum value (and similar problems).

Following this work, a word about computational issues is in place. For the approximation results by homogeneous markets, one can easily find the best homogeneous market by computing the revenue of each one of the $n$ possible homogeneous markets (assuming a suitable computational model that supports computing the revenue of a homogeneous market). Similarly, the existential $(1-\eps)$-approximation result can be made concrete by checking each of the $poly(n^{f(\eps)})$ economies that use at most $f(\eps)$ distributions. However, if possible, it will be interesting to develop algorithms that find an economy for which 
the second price auction extracts a $(1-\eps)$-fraction of the ideal second price revenue in time $poly(n,\frac 1 \eps)$.
\else
    The full paper can be found at: \url{https://arxiv.org/xxxxx}
\fi

\iffull
	\section{Optimal Auctions in Homogeneous Economies}\label{sec:optimal_auction}

In this section we prove our main result:

\begin{theorem}
Let $D=\{\mathcal D_1,\ldots, \mathcal D_n\}$ be a set of distributions. There exists a homogeneous economy $\mathbf E$ where the value of each bidder $i$ is independently drawn from the same distribution $\mathcal D\in D$, such that the revenue that can be generated by an optimal auction for $\mathbf E$ is at least a $\left (\frac 1 2 \cdot (1-\frac 1 e)\right )$-fraction of the ideal revenue.
\end{theorem}

We note that the auction that we run in the economy $\mathbf E$ is very simple: a second-price auction with an appropriately chosen reserve price. 

For the proof, let the optimal economy be $\Eopt\in \arg\max_\econ\rev$. For $i\in[n]$, let $\D_i$ be the distribution of the value of bidder $i$ in $\Eopt$. 
We split the probability mass of each distribution $\D_i$ into low and high values using a threshold $H$ derived as follows: for each distribution $\D_i$, let $h_i$ be the infimum of the values $T$ that satisfy $\Pr_{X\sim \D_i}[X\leq T]\geq 1- \frac1n$. Define the threshold $H$ as
$$
	H=\max_{i\in[n]} h_i
$$
Notice that $\Pr_{X\sim \D_i}[X > H]\leq \frac1n$, for all $i\in[n]$. Values that are at most $H$ are considered \emph{low values}, while all other values are considered \emph{high values}.

For every economy $\econ$, define $\lowrev$ to be the expected revenue (or contribution) given that the value of the winner is low. Define $\highrev$ similarly to be the expected revenue given that the value of the winner is high.
Clearly
$$
	\rev[\Eopt]= \lowrev[\Eopt]+\highrev[\Eopt]
$$
Define $\rev[\Eopt]=\opt$ as the ideal revenue. Therefore, either the low values or the high values must contribute at least $\frac12\opt$. In each case, we show that a second price auction with a reserve price in some homogenous economy obtains revenue of at least $\frac12\cdot(1-\frac1e)\opt$.

\paragraph{Case I: Low values contribute at least $\boldsymbol{\frac12\cdot \opt}$.} Notice that the revenue from any low-value winning bidder is at most $H$. Hence, in the current case
\begin{equation}\label{eq:H>OPT/2}
  	H \geq \frac12\cdot\opt
\end{equation}
By the definition of $H$, there exists a distribution $\D_H$ from $\Eopt$ such that $\Pr_{X\sim \D_H}\big[X < H\big]\leq1-\frac1n$. 
Consider the homogenous economy, denoted $\econ'$, consisting of $n$ bidders taken from distribution $\D_H$. Denote the values of those bidders $X_1,\ldots,X_n \sim \D_H$. We now obtain:
$$
	\Pr\big[\maxone(X_1,\ldots, X_n) \geq H\big]%
		= 1-\prod_{i=1}^n\Pr\big[X_i< H\big]%
		\geq 1-\left(1-\frac1n\right)^n \geq 1-\frac1e
$$
where the equality is due to bidder values being i.i.d., the first inequality is by the definition of $\D_H$, and the second inequality follows from the fact that $(1-\frac1n)^n\leq \frac1e$, for any natural number $n$.

Therefore, the revenue of a second price auction with reserve price $H$ is at least $(1-\frac 1 e)\cdot H$ (the highest value exceeds the reserve price with probability $(1-\frac 1 e)$). Hence, by Equation~\eqref{eq:H>OPT/2} we obtain
$$
	\rev[\econ']\geq \left(1-\frac1e\right)\cdot H 
				\geq \frac12 \cdot \left(1-\frac1e\right)\opt		
$$

\paragraph{Case II: High values contribute at least $\boldsymbol{\frac12\cdot \opt}$.} In this case we focus on instances where the winning bid is a high value and ignore all contribution from low values. Thus, we round down the low values of each distribution to zero, and keep only the probability mass of values exceeding $H$. Let $\tD_1,\ldots,\tD_n$ denote these ``rounded down'' distributions of the bidders from $\Eopt$. Denote the corresponding economy $\econtilde$. Clearly,\footnote{Notice that modifying the optimal auction over $\Eopt$ by changing the price to $H$ whenever it is lower than $H$ we lose none of the contribution of high values.}
\begin{equation}\label{eq:high_rev>OPT/2}
  	\rev[\econtilde]\geq \frac12 \opt
\end{equation}
Define $r_i= \expct\big[\text{Revenue gained from bidder $i$ in the optimal auction over $\econtilde$}\big]$.
Without loss of generality, assume $r_1=\max_i r_i$. Thus,
\begin{equation}\label{eq:rev<n*r_1}
\rev[\econtilde]=\sum_{i=1}^n r_i \leq n\cdot r_1
\end{equation}
Consider an optimal auction with one bidder from $\tD_1$. Note that the optimal auction for a single bidder is a take-it-or-leave-it auction \cite{M81}. Let $p_1$ be the take-it-or-leave-it offer of this auction, and $q_1$ be the probability that the bidder accepts. The expected revenue is equal to $p_1\cdot q_1$. Moreover, notice that the revenue of this single-bidder optimal auction is at least $r_1$. That is,
\begin{equation}\label{eq:r_1<p_1*q_1}
  	r_1 \leq p_1\cdot q_1
\end{equation}
Therefore, plugging Equation~\eqref{eq:r_1<p_1*q_1} into Equations~\eqref{eq:high_rev>OPT/2} and~\eqref{eq:rev<n*r_1}, we obtain
\begin{equation}\label{eq:n*p_1*q_1>OPT/2}
  	n\cdot (p_1\cdot q_1) \geq n\cdot r_1 \geq \rev[\econtilde]\geq \frac12\opt
\end{equation}
Moreover, notice that 
\begin{equation}\label{eq:q_1<1/n}
	q_1 = \Pr_{X\sim \tD_1}\big[X\geq p_1\big] \leq \Pr_{X\sim\tD_1}\big[X > 0\big] \leq \frac1n~  
\end{equation}
where the equality is by definition of $q_1$,  the first inequality is due to $p_1>0$ (otherwise, $\opt=0$ by Equation~\eqref{eq:n*p_1*q_1>OPT/2}) and the second inequality follows from the fact that all the probability mass at most $H$ has been shifted to zero in the distribution $\tD_1$, and thus the remaining positive mass in $\tD_1$ is at most $\frac1n$ by the definition of $H$. 

To conclude the proof we require the following known claim. Proof included in Appendix \ref{apx:missing-proofs}.
\begin{claim}\label{clm:1-(1-x)^n-bound}
		$1-(1-x)^n \geq (1-\left(1-\frac1n\right)^n)\cdot nx$ for every $x\in\big[0,\frac1n\big]$.
\end{claim}
\long\def\CLM22proof{
\noindent{\bf Proof of Claim \ref{clm:1-(1-x)^n-bound}:~~}
	The claim is trivial for $n=1$. Thus, assume $n\geq2$. Let $$f(x)=1-(1-x)^n \qquad \mbox{and}\qquad g(x)=\left(1-\left(1-\frac1n\right)^n\right)\cdot nx$$
	Differentiating $f$ twice yields $f''(x)=-n(n-1)(1-x)^{n-2}$, which is negative for all $x\in[0,1/n]$. Thus, $f$ is concave in this range. Notice that $g(x)$ is a linear function of $x$ and that $f(0)=g(0)$ and $f(1/n)=g(1/n)$. Therefore, by concaveness, we have that
	$f(x)\geq g(x)$
for any $x\in[0,1/n]$.
\qed
}

Finally, let $\econ_1$ and $\econtilde_1$ be homogeneous economies consisting of $n$ bidders drawn from distributions $\D_1$ and $\tD_1$, respectively. Let $SP$ be the revenue of the second price auction over economy $\econtilde_1$ with reserve price $p_1$, and let $X_i\sim \tD_1$ denote the value of bidder $i$. Clearly, $\rev[\econ_1]\geq \rev[\econtilde_1]\geq SP$. Thus, we obtain
\begin{align*}
	\rev[\econ_1]\geq SP \;\geq&\;\; p_1 \cdot \Pr\big[\exists i\in [n]: X_i \geq p_1\big] & \nonumber \text{(we obtain $\geq p_1$ if any bid exceeds $p_1$)}\\[1em]
	\geq&\;\;  p_1\cdot\left(1-\Pr\big[\forall i \in [n]: X_i< p_1\big]\right) & \nonumber\\
	\geq&\;\;  p_1\cdot\Big(1-\prod_{i=1}^{n}\Pr[X_i< p_1]\Big) & \text{(since $X_1,\ldots,X_n$ are i.i.d.)} \nonumber\\
	\geq&\;\;  p_1\cdot\left(1-\left(1-q_1\right)^n\right) & \text{(by definition of $q_1$)}\nonumber\\
	\geq&\;\;  p_1\cdot\left(1-\left(1-\frac1n\right)^n\right)\cdot n \cdot q_1 & \text{(by Equation~\eqref{eq:q_1<1/n} and Claim~\ref{clm:1-(1-x)^n-bound})} \nonumber\\
	\geq&\;\;  p_1\cdot\left(1-\frac1e\right)\cdot n \cdot q_1 & \text{(as $\left(1-\frac1n\right)^n\leq \frac1e$ )} \nonumber\\
 	\geq&\;\; \frac12\cdot\left(1-\frac1e\right)\cdot\opt &  \text{(by Equation~\eqref{eq:n*p_1*q_1>OPT/2})}
\end{align*}

	\section{Second Price Auctions in Homogeneous Economies}

We now prove that there exists a homogeneous economy whose revenue for a second price auction is at least a $c$-fraction of the ideal second price revenue, for some constant $c>0$. Formally:

\begin{theorem}
Let $D=\{\mathcal D_1,\ldots, \mathcal D_n\}$ be a set of distributions. There exists a homogeneous economy $\mathbf E$, in which each bidder's value is independently drawn from the same distribution $\mathcal D\in D$, for which the revenue obtained by a second price auction in $\mathbf E$ is at least a $c$-fraction of the ideal second-price revenue, for some constant $c>0$.
\end{theorem}

In Section~\ref{sec:ptas} we use the techniques introduced in the proof of this theorem to show that the ideal second-price revenue can be approximated arbitrarily well by choosing only a few distributions from $D$. Thus, in addition to establishing one of our main results, this section also serves as a warm-up section to the more technically involved construction of Section \ref{sec:ptas}.

Let $\opt$ be the optimal second price revenue, i.e., $\opt=\max_\econ{\sprice}$. Let $\Eopt$ be an economy whose second price revenue is $\opt$. Let $\D_i$ be the distribution of the value of bidder $i$ in $\Eopt$, for $i\in[n]$. Similarly to Section~\ref{sec:optimal_auction}, we partition the probability mass of each distribution into low and high values.
Define the threshold $H$ as
$$
	H=\max_i\inf\left\{h_i\;\Bigg| \;\Pr_{X\sim \D_i}[X\leq h_i]\geq 1-\frac{1}{n-1}\right\}
$$
Notice that $\Pr_{X\sim \D_i}\big[X> H\big]\leq \frac1{n-1}$, for all $i\in[n]$,  and that there must be some $j\in[n]$ (namely, the maximizer of the term above), such that $\Pr_{X\sim \D_j}\big[X\geq H\big]\geq \frac1{n-1}$.
We refer to auctions where the second-highest value is at most $H$ as \emph{low-value auctions} and to auctions in which the second-highest value is at least $H$ as \emph{high-value auctions}. Clearly, the sum of the contribution of low values and the contribution of high values is equal to the expected second-highest value, i.e., $\opt$. Hence, we divide into two cases: First, when low values  contribute at least $\alpha\cdot\opt$ to the optimal revenue, and second, when high values contribute at least $(1-\alpha)\cdot\opt$, for some constant $\alpha\in[0,1]$ to be determined later.

\subsection{Case I: Low Values Contribute at Least $
\alpha\cdot\opt$}

The second highest value in each low-value auction is at most $H$. Hence, the contribution of low values is at most $H$. Therefore
\begin{equation}\label{eq:H>alpha*OPT}
  	H\geq \alpha\cdot \opt
\end{equation}
Before the next step of our analysis, we require the following simple claim.
\begin{claim}\label{clm:x(1-x)^(n-1)_non-increasing}
	For every natural $n\geq2$, the function $f(x)=x(1-x)^{n-1}$ is non-increasing for $x\in[\frac{1}{n},1]$.
\end{claim}
\begin{proof}
	Differentiating $f$ yields $f'(x)=(1-x)^{n-2}(1-nx)$, which is non-positive for $x\in[\frac{1}{n},1]$.
\end{proof}

By the definition of $H$, there exists a distribution from $\Eopt$ such that $\Pr_{X\sim \D_H}\big[X \geq H\big]\geq\frac{1}{n-1}$. Thus by Claim~\ref{clm:x(1-x)^(n-1)_non-increasing}, it follows that
\begin{equation}\label{eq:D_H_probs_bound}
  	\Pr_{X\sim \D_H}\big[X \geq H\big]\cdot \left(\Pr_{X\sim \D_H}\big[X < H\big]\right)^{n-1}\leq \frac{1}{n-1}\left(1-\frac1{n-1}\right)^{n-1}
\end{equation}
Consider the homogenous economy, denoted $\econ'$, consisting of $n$ bidders taken from distribution $\D_H$.
Let $X_1,\ldots,X_n \sim \D_H$ be the random variables that denote the values of these bidders. We use the notation $\smax$ to denote the second largest value from a given set of values. We now obtain 
\begin{align}
	\Pr\big[\smax(X_1,\ldots, X_n) \geq H\big] \;=&\;\; \Pr[\mbox{\emph{at least two} bidders exceed $H$}]  \nonumber \\[1ex]
	 \;=&\;\; 1- \Pr[\mbox{\emph{at most one} bidder exceeds $H$}] \nonumber \\
	=&\;\;  1- \prod_{i=1}^n\Pr\big[X_i < H\big]-\sum_{j=1}^n\Pr\big[X_i\geq H\big]\prod_{j\neq i}\Pr\big[X_j < H\big]  \nonumber\\
	\geq&\;\;  1-\left(1-\frac1{n-1}\right)^n-n\cdot\frac1{n-1}\left(1-\frac1{n-1}\right)^{n-1}  \nonumber\\
	=&\;\;  1-\left(1-\frac1{n-1}\right)\left(1-\frac1{n-1}\right)^{n-1}-\left(1+\frac1{n-1}\right)\left(1-\frac1{n-1}\right)^{n-1}  \nonumber\\
	\geq&\;\;  1-2\cdot\frac1{e}~,\label{eq:prob>1-2/e}  
\end{align}
where the third equality is due to bidder values being i.i.d., the first inequality is by Equation~\eqref{eq:D_H_probs_bound}, and the second inequality follows from the fact that $(1-\frac1n)^n\leq \frac1e$, for any natural number $n$. Therefore, by Equations~\eqref{eq:H>alpha*OPT} and~\eqref{eq:prob>1-2/e} we conclude that:
$$
	\sprice[\econ']=\expct\big[\smax(X_1,\ldots, X_n)\big]\geq H\cdot\Pr\big[\smax(X_1,\ldots, X_n) \geq H\big]\geq \alpha\cdot\left(1-\frac2e\right)\cdot\opt
$$

\subsection{Case II: High Values Contribute at Least $(1-\alpha)\cdot\opt$}

Here we focus on auctions where the second-highest value is a high value. We may therefore ignore all contribution from low values. Thus, we round down the low values of each distribution to zero, and keep only the probability mass of values exceeding $H$. Let $\tD_1,\ldots,\tD_n$ denote these ``rounded down'' distributions of the bidders from $\Eopt$. Denote the corresponding economy $\econtilde$. Clearly,
\begin{equation}\label{eq:high_rev>(1-alpha)OPT}
  	\sprice[\econtilde]= \expct_{V_i\sim\tD_i}\big[\smax(V_1,\ldots,V_n)\big]\geq (1-\alpha) \opt
\end{equation}

In the proof of this case, we first show that there exist two distributions $\D_i$ and $\D_j$ such that if the value of every bidder is drawn from one of these distributions then the expected revenue of a second-price auction is a constant fraction $c>0$ of $\opt$. Then, we show that for some $\D\in\{\D_i,\D_j\}$ the homogeneous economy where the value of each bidder is independently drawn from $\D$ also generates a $c$-fraction of $\opt$.

In the remainder of this section, we use $V_i$ to denote a random variable that is drawn from distribution $\tD_i$.


\subsubsection{The Revenue of Economies with Two Distributions}

We consider the contribution of every possible unordered
pair of distributions, by restricting our focus to instances where a specific pair of bidders $i$ and $j$, have drawn the two largest values.
Let $B_{i,j}$ denote the event where the values of bidders $i$ and $j$ are the two highest values, for each $1\leq i<j\leq n$. Let $r_{i,j}$ be the contribution of such instances:
$$
	r_{i,j}=\expct
	\big[\smax(V_1,\ldots,V_n)\cdot\one_{B_{i,j}}\big]
$$
Without loss of generality assume $r_{1,2}=\max_{1\leq i<j\leq n}{r_{i,j}}$.
Notice that the events $B_{i,j}$ naturally partition the instances of the economy $\econtilde$ to disjoint sets, since for a given instance only one event $B_{i,j}$ can happen. Thus the revenue of a second-price auction can be bounded in terms of $r_{1,2}$:
\begin{equation}\label{eq:sp<n^2*r_12}
  \expct
  \big[\smax(V_1,\ldots,V_n)\big] \leq \sum_{i<j}{r_{i,j}}\leq {n \choose 2}\cdot r_{1,2}~
\end{equation}
where the second inequality is due to the maximality of $r_{1,2}$. Moreover, notice that given the event $B_{i,j}$ the random variable $\smax(V_1,\ldots,V_n)$ is equal to the random variable $\smax(V_i,V_j) $. Thus,
\begin{equation}\label{eq:r_i,j<smax(D_i,D_j)}
  r_{i,j}=\expct
  			\big[\smax(V_1,\ldots,V_n)\cdot\one_{B_{i,j}}\big]
  		 =\expct
  		 	\big[\smax(V_i,V_j)\cdot\one_{B_{i,j}}\big]
  		 \leq \expct
  		 	\big[\smax(V_i,V_j)\big]
\end{equation}
Let $X_1,\ldots,X_n\sim\tD_1$ be random variables drawn from $\tD_1$ and $Y_1,\ldots,Y_n\sim\tD_2$ be random variables drawn from $\tD_2$. Consider the following three economies: $\econ_1$ and $\econ_2$ are the homogeneous economies where the values of the bidders are represented by the random variables $X_1,\ldots,X_n$ and $Y_1,\ldots,Y_n$, respectively, and $\econ_{12}$ is the two-distribution economy consisting of bidder values $X_1,\ldots,X_{n/2},Y_{n/2+1},\ldots,Y_n$,
i.e., half of the bidder values are drawn from each of the two distributions $\tD_1$ and $\tD_2$. We focus on $\econ_{12}$ first, showing its revenue approximates $\opt$. We then show that either $\econ_1$ or $\econ_2$ achieves the same approximation factor as $\econ_{12}$.

The following random variables shall be useful when comparing the three economies:
\begin{align*}
	\Xonemax&=\maxone(X_1,\ldots,X_{\frac{n}2}); &
	\Xtwomax&=\maxone(X_{\frac{n}2+1},\ldots,X_n); 	\\
	\Yonemax&=\maxone(Y_1,\ldots,Y_{\frac{n}2}); &
	\Ytwomax&=\maxone(Y_{\frac{n}2+1},\ldots,Y_n). 
\end{align*}
Note that we always have $\smax(X_1,\ldots,X_{\frac{n}2},Y_{\frac{n}2+1},\ldots,Y_n)\geq \min\{\Xonemax,\Ytwomax\}$. Hence:
\begin{equation}\label{eq:smax>min(X_max,Y_max)}
	\sprice[\econ_{12}]=\expct\big[\smax(X_1,\ldots,X_{\frac{n}2},Y_{\frac{n}2+1},\ldots,Y_n)\big]
						\geq \expct\big[\min\{\Xonemax,\Ytwomax\}\big]
\end{equation}
Let $Z_{i,j}$, for $1\leq i<j\leq n$, denote the event where all bidders aside from bidder $i$ and $j$ have zero value. Recall that we discarded all the probability mass at most $H$, so by the definition of $H$ the probability that the value of a specific bidder is $0$ is at least $1-\frac1{n-1}$. Hence, for every $1\leq i<j\leq n$,
\begin{equation}\label{eq:Z_ij}
  \Pr[Z_{i,j}]\geq \left(1-\frac1{n-1}\right)^{n-2}
  				=\;\frac{1}{\left(1+\frac{1}{n-2}\right)^{n-2}}  
  				\;\geq\; \frac1e
\end{equation}
We thus obtain:
\begin{align*}
	\expct\big[\min\{\Xonemax,\Ytwomax\}\big] \;\geq&\;\; 
	\expct\bigg[\min\{\Xonemax,\Ytwomax\}\cdot\one\bigg\{\bigvee_{i=1}^{n/2}\bigvee_{j=n/2+1}^nZ_{i,j}\bigg\}\bigg]  &\nonumber \\
	 \;=&\;\; \sum_{i=1}^{n/2}\sum_{j=n/2+1}^n\expct\left[\min\{\Xonemax,\Ytwomax\}\cdot\one_{Z_{i,j}}\right]  &\text{($Z_{i,j}$ are disjoint events)}\nonumber \\
	=&\;\; \sum_{i=1}^{n/2}\sum_{j=n/2+1}^n\expct\left[\min\{X_i,Y_j\}\cdot\one_{Z_{i,j}}\right]  &\text{(by the definition of $Z_{i,j}$)}\   \nonumber\\
	=&\;\; \sum_{i=1}^{n/2}\sum_{j=n/2+1}^n\expct\left[\min\{X_i,Y_j\}\right]\cdot\Pr[Z_{i,j}]   &\text{(independent r.v.'s)}\nonumber\\
	=&\;\; \sum_{i=1}^{n/2}\sum_{j=n/2+1}^n\expct\left[\min\{V_1,V_2\}\right]\cdot\Pr[Z_{i,j}]	&\text{($X_i\sim \tD_1$ and $Y_j \sim \tD_2$)}\\
	\geq&\;\; \left(\frac{n}2\right)^2\cdot r_{1,2}\cdot\frac1e &\text{(by Eq.~\eqref{eq:r_i,j<smax(D_i,D_j)}
	 and~\eqref{eq:Z_ij})}  \nonumber \\
	\geq&\;\; \frac12\cdot(1-\alpha)\opt\cdot\frac1e &\text{(by Eq.~\eqref{eq:high_rev>(1-alpha)OPT} and~\eqref{eq:sp<n^2*r_12})}
\end{align*}
To summarize, we obtain
\begin{equation}\label{eq:min(X_max,Y_max)>(1-alpha)/(2e)OPT}
  \expct\big[\min\{\Xonemax,\Ytwomax\}\big]\geq \frac{1-\alpha}{2e}\cdot\opt
\end{equation}

\subsubsection{The Revenue of a Homogeneous Economy}

Next we show that the bound $\expct\big[\min\{\Xonemax,\Ytwomax\}\big]$, proved on the revenue of a second price auction in the economy $\econ_{12}$, in fact holds for one of the homogeneous economies $\econ_1,\econ_2$. We shall require the following simple claim.
\begin{claim}\label{clm:xy<0.5(x^2+y^2)}
	For every $x,y\in\RR$, $xy\leq(x^2+y^2)/2$.
\end{claim}
\begin{proof}
	Since squares of real numbers are non-negative, we get $(x-y)^2\geq 0$. Thus, $x^2-2xy+y^2\geq0$. Rearranging, we obtain the claim.	
\end{proof}

As $\min\{\Xonemax,\Ytwomax\}$ is a non-negative random variable, we can write its expectation as:
\begin{align}
	\expct\big[\min\{\Xonemax,\Ytwomax\}\,\big] \;=&\;\; 
	\int_0^\infty\Pr\big[\min\{\Xonemax,\Ytwomax\}\geq t\,\big]dt \nonumber  \\
	=&\;\; \int_0^\infty\Pr\big[(\Xonemax\geq t)\wedge(\Ytwomax\geq t)\big]dt \nonumber \\
	=&\;\; \int_0^\infty\Pr\big[\Xonemax\geq t\big]\cdot\Pr\big[\Ytwomax\geq t\big]dt  \nonumber \\
	\leq&\;\; \frac12\int_0^\infty\left(\Pr\big[\Xonemax\geq t\big]\right)^2dt+\frac12\int_0^\infty\left(\Pr\big[\Ytwomax\geq t\big]\right)^2dt\label{eq:UB_for_min(X_max,Y_max)}
\end{align}
where the last inequality follows from Claim~\ref{clm:xy<0.5(x^2+y^2)}.
Note that we always have $\smax(X_1,\ldots,X_n)\geq \min\{\Xonemax,\Xtwomax\}$. Therefore,
\begin{equation}\label{eq:smax>min(X_max,X_max)}
	\sprice[\econ_{1}]=\expct\big[\smax(X_1,\ldots,X_n)\big]
						\geq \expct\big[\min\{\Xonemax,\Xtwomax\}\,\big]
\end{equation}
Analogously to our calculations above for $\expct\big[\min\{\Xonemax,\Ytwomax\}\,\big]$, we obtain
\begin{equation}\label{eq:min(Xmax,Xmax)=integral}
  \expct\big[\min\{\Xonemax,\Xtwomax\}\,\big] \;=\; 
	\int_0^\infty\Pr\big[\Xonemax\geq t\big]\cdot\Pr\big[\Xtwomax\geq t\big]dt=
	\int_0^\infty\left(\Pr\big[\Xonemax\geq t\big]\right)^2dt ~
\end{equation}
where the last equality is due to  $\Xonemax$ and $\Xtwomax$ being i.i.d. Without loss of generality assume
$$\int_0^\infty\left(\Pr\big[\Xonemax\geq t\big]\right)^2dt \geq \int_0^\infty\left(\Pr\big[\Ytwomax\geq t\big]\right)^2dt~$$
(the proof is analogous for the other case). Therefore, by Equations~\eqref{eq:min(X_max,Y_max)>(1-alpha)/(2e)OPT},\eqref{eq:UB_for_min(X_max,Y_max)},\eqref{eq:smax>min(X_max,X_max)} and~\eqref{eq:min(Xmax,Xmax)=integral}:
$$
	\sprice[\econ_1]\geq  \expct\big[\min\{\Xonemax,\Xtwomax\}\,\big]
					\geq  \expct\big[\min\{\Xonemax,\Ytwomax\}\,\big]
					\geq  \frac{1-\alpha}{2e}\cdot\opt
$$
This concludes our proof for Case II.

\subsection{Choosing the Value of $\alpha$}
In Case I we showed that the revenue of a second price auction in a homogeneous economy is at least $\left(1-\frac2e\right)\alpha\cdot\opt$ whereas in Case II we proved a bound of $\frac{1-\alpha}{2e}\cdot\opt$. By choosing $\alpha = \frac{1}{2e-3}$ we maximize the lower of these two bounds, showing that there is always a homogeneous economy where the revenue of a second price auction is at least $\left(1-\frac2e\right)\cdot\frac{1}{2e-3}\cdot \opt\approx 0.10844\cdot \opt$.

\section{$\boldsymbol{(1-\eps)}$-Approximation for the Second Price Auction}\label{sec:ptas}

In the previous sections we showed that there exists one homogeneous economy that provides a constant approximation to the ideal revenue (or to the ideal second-price revenue). In this section we show that we can get arbitrarily close to the ideal second-price revenue using only relatively few distributions:
\begin{theorem}
Let $D=\{\mathcal D_1,\ldots, \mathcal D_n\}$ be a set of 
distributions. For every constant $\eps>0$ there exists a subset of distributions $D'\subseteq D$, $|D'|\leq (1/\eps)^{O(1/\eps^{10})}$, and an economy $\Eeps$ where the value of each bidder is independently drawn from some distribution $\mathcal D\in D'$, for which the revenue obtained by a second price auction in $\Eeps$ is at least a $\left (1-\eps \right )$-fraction of the ideal second-price revenue.
\end{theorem}
In other words, for every constant $\eps>0$, a constant number of distributions is needed to construct an economy that guarantees a $(1-\eps)$ of the revenue of the ideal second-price revenue. 


Denote the optimal economy $\Eopt$, and,  for $i\in[n]$, let $\D_i$ be the distribution of bidder $i$ in economy $\Eopt$. We show that:
$$
	\sprice[\Eeps]\geq (1-\eps)\sprice[\Eopt]
$$

Roughly speaking, the proof partitions the supports of the distributions into three main segments: low, middle, and high. Similarly to previous sections, the revenue that comes from low values is relatively easy to handle by having enough bidders whose value is drawn from a distribution in which the largest low value is obtained with high enough probability. To handle the revenue due to middle values, we group together several distributions based on the similarity of their middle quantiles. The grouping process is technically subtle, but we manage to show that the distributions can be grouped into a relatively small number of groups such that the distributions within each group are almost completely ``exchangeable'' in terms of the contribution of their middle values to the revenue. The challenging part of the proof is to handle the revenue due to high values. The issue is that it is not possible to partition the distribution into a small number of groups since the expected contribution of the highest values cannot be properly bounded. We therefore develop a carefully constructed sampling method that allows us to prove that a small number of distributions can provide almost all of the revenue due to high values.

For simplicity of presentation, in the remainder of this proof we assume that all distributions in $D$ are atomless\footnote{This is in fact without loss of generality since a second-price auction is ``smooth'' in the following sense: if we replace a distribution $\mathcal D$ of a bidder in a second price auction by slightly perturbing it, the revenue of the second price auction will change only slightly. In particular, we can take any distribution $\mathcal D$ with atoms and ``smooth'' it by replacing each value $v$ that we get with probability $p>0$ by a uniform distribution on $[v,v+\eps]$ with total mass $p$.}.

\subsection{Defining Low, Middle, and High Values}

As in previous sections, we partition the probability mass of each distribution. In this section, however, we partition into low, middle and high values (rather than just low and high values).
Define the thresholds $L$ and $H$ as:
$$
	L=
	\max_i\inf\left\{\ell_i\;\Bigg| \;\Pr_{V_i\sim\D_i}[V_i\leq \ell_i]\geq 1-\frac{1}{\eps^2n}\right\};
\quad\quad
	H=
	\max_i\inf\left\{h_i\;\Bigg| \;\Pr_{V_i\sim\D_i}[V_i\leq h_i]\geq 1-\frac{\eps}{n}\right\}.
$$
For each distribution, we partition the probability mass as follows. The probability mass in $[0,L)$ is regarded as \emph{low values}. The probability mass in $[L,H)$ is regarded as \emph{middle values}. The probability mass in $[H,\infty)$ is regarded as \emph{high values}. For an economy $\econ$, with bidder values $X_1,\ldots,X_n$, define:
\begin{align*}
\lowsprice&=\expct\big[\smax(X_1,\ldots,X_n)\one\{\smax(X_1,\ldots,X_n)<L\}\big];\\
\midsprice&=\expct\big[\smax(X_1,\ldots,X_n)\one\{\smax(X_1,\ldots,X_n)\in[L,H)\}\big]; \\
\highsprice&=\expct\big[\smax(X_1,\ldots,X_n)\one\{\smax(X_1,\ldots,X_n)\geq H\}\big].
\end{align*}
Clearly,
$$ \sprice=\lowsprice+\midsprice+\highsprice$$
The economy $\Eeps$ that we construct will satisfy:
\begin{equation*}
 \lowsprice[\Eeps]\geq (1-\eps)\cdot\lowsprice[\Eopt] \qquad\;
	\midsprice[\Eeps]\geq (1-\eps)\cdot\midsprice[\Eopt]  \qquad\;
	\highsprice[\Eeps]\geq (1-\eps)\cdot\highsprice[\Eopt]
\end{equation*}
which would imply what we need:
$$
	\sprice[\Eeps]\geq (1-\eps)\sprice[\Eopt]
$$

\subsection{Approximating the Low Values}
By the definition of $L$, there exists a distribution $\D_L$ such that $\Pr_{X\sim\D_L}[X\geq L]=1/(\eps^2n)$.
\begin{lemma}\label{lem:low_value_approx}
	The second highest value among $\eps\cdot n$ bidders with values drawn i.i.d. from $\D_L$ is at least $L$ with probability $1-3\eps$.
\end{lemma}
\begin{proof}
Let $X_1,\ldots,X_{\eps n}\sim\D_L$. It suffices to show that 
\begin{equation}\label{eq:prob_low_UB}
  	\Pr[\smax(X_1,\ldots,X_{\eps n})<L] \leq 3\eps
\end{equation}
Notice that $\Pr[X_i\geq L]=1/(\eps^2 n)$, for all $i\in[\eps n]$, by definition of $\D_L$, and thus
\begin{align}
	\Pr\big[\smax(X_1,\ldots, X_{\eps n}) < L\big] \;=&\;\; \Pr[\mbox{\emph{at most one} bidder exceeds $L$}] \nonumber \\
	=&\;\;\prod_{i=1}^{\eps n}\Pr\big[X_i < L\big]+\sum_{j=1}^{\eps n}\Pr\big[X_i\geq L\big]\prod_{j\neq i}\Pr\big[X_j < L\big]  \nonumber\\
	=&\;\;  \left(1-\frac1{\eps^2n}\right)^{\eps n}+\eps n\cdot\frac1{\eps^2n}\left(1-\frac1{\eps^2n}\right)^{\eps n-1}  \nonumber\\
	=&\;\;  \left(1+\frac1{\eps} \left(1-\frac1{\eps^2n}\right)^{-1}\right)\left(1-\frac1{\eps^2n}\right)^{\eps n} \label{eq:prob_low_is_small}
\end{align}
We may assume that $\left(1-\frac1{\eps^2n}\right)^{-1}<2$ (otherwise, $n\leq 1/\eps^2$ and thus setting $\Eeps=\Eopt$ would be sufficient as our solution). Moreover, $1\leq\frac1\eps$ and $\left(1-\frac1{\eps^2n}\right)^{\eps n}\leq e^{-\frac1\eps}$. Plugging these three facts into Equation~\eqref{eq:prob_low_is_small}, by Equation~\eqref{eq:prob_low_UB} it suffices to show
$\frac3\eps\cdot e^{-\frac1\eps}\leq 3\eps$.

Rearranging, we obtain $\frac1{2\eps}-\ln\frac1\eps\geq0$. The function $\frac{x}2-\ln{x}$ is convex and attains its unique minimum at $x=2$. Moreover, $1-\ln2>0$, so the minimum value is positive. It follows that the inequality above holds for all $\eps>0$. This proves the lemma.
\end{proof}

Lemma~\ref{lem:low_value_approx} provide an approximation of $1-3\eps$ to the revenue of $\Eopt$, restricted to instances where the revenue is at most $L$. The next lemma shows that having ``sacrificed'' $\eps n$ bidders, we can still get very close to $\opt$ by choosing some set of $(1-\eps)n$ bidders from $\Eopt$.

\begin{lemma}\label{lem:close_to_OPT_without_eps*n_bidders}
		For any economy $\econ$, and for any $\eps>0$, there exists an economy $\econ'$, consisting of a subset of size $(1-\eps)n$ of the bidders from $\econ$, such that 
		$$
			\sprice[\econ']\geq (1-2\eps)\sprice
		$$
\end{lemma}
\begin{proof}
	Let $X_1,\ldots,X_n$ denote the values of the bidders in economy $\econ$. Let $S\subseteq[n]$ be a set of $\eps n$ bidders chosen uniformly at random. We call the bidders in $S$ \emph{virtual} bidders. We run a second price auction in economy $\econ$, except that if one of the bidders in $S$ is one of the two bidders with the highest value, the item is not allocated at all. We show that the expected revenue of this auction is high, which shows that a second price auction in the economy  $\econ'$, which is the economy consisting of all non-virtual bidders from $\econ$, $[n]\setminus V$, is at least as high.
	
Each specific bidder has probability $(1-\eps)$ of not being in $S$, and two specific bidders are both non-virtual with probability $(1-\eps)^2\geq 1-2\eps$. Therefore,
$$
	\expct_V\big[\sprice[\econ']\big]\geq (1-2\eps)\sprice
$$
This holds in expectation, thus there is a specific choice of $S$ satisfying it. The lemma follows.
\end{proof}

By Lemma~\ref{lem:close_to_OPT_without_eps*n_bidders} there exists an economy $\Eopt'$ with $(1-\eps)n$ bidders that attains $(1-2\eps)\opt$.
For notational convenience we keep referring to $\Eopt$ and refrain
from ``shrinking'' it to $\Eopt'$ at this point. We consider the low values taken care of, but we only apply Lemmata~\ref{lem:low_value_approx} and~\ref{lem:close_to_OPT_without_eps*n_bidders} to add the low value bidders at the end, after all other bidders of $\Eeps$ have been decided

\subsection{Approximating the Middle Values}
Our approach for middle values is to keep them roughly as they were
in $\Eopt$ when constructing $\Eeps$. For each bidder $i$ her distribution over the middle values in $\Eopt$ would be similar to that in $\Eeps$. To this end we ``discretize'' the range of middle values, $[L,H)$ in each distribution $\D_i$, and divide the distributions into groups.
A pair of distributions belongs to the same group if their ``discretized''
middle values are the same.

We first round each middle value down to the nearest multiple of $\eps\opt$. Let $L=v_0,v_1,\ldots,v_m=H$ be those multiples, i.e.,
\begin{equation}\label{eq:defn_of_v_i}
  	v_i=L+ i\cdot\eps\opt
\end{equation}
for each $i\in[m]$. That is, for each distribution $\D_i$ in $\Eopt$, define the rounded distribution $\D_i'$ as the discrete distribution that attains the value $v_j$ with probability $p_{i,j}=\Pr_{V_i\sim\D_i}\big[V_i\in[v_i,v_{i+1})\big]$ for $i\in[m-1]$, and the value $v_m=H$ with probability $p_{i,m}=\Pr_{V_i\sim\D_i}\big[V_i\geq H\big]$.  Next, observe that the number of such multiples of $\eps\opt$ depends only on $\eps$, as the following lemma implies.

\begin{lemma}\label{lem:UB_of_H}
		$H\leq 4\opt/\eps^3$.
\end{lemma}
\begin{proof}
	Assume towards contradiction that $H>4\opt/\eps^3$. By the definition of $H$, there exists a distribution $\D_H$ in $\Eopt$ such that $\Pr_{X\sim\D_H}[X\geq H]=\eps/n$. Let $X_1,\ldots,X_n\sim\D_H$ be $n$ values drawn from $\D_H$. We obtain,
\begin{align*}
		\Pr\big[\smax(X_1,\ldots,X_n)\geq H\big]&=1- \prod_{i=1}^n\Pr\big[X_i < H\big]-\sum_{j=1}^n\Pr\big[X_i\geq H\big]\prod_{j\neq i}\Pr\big[X_j < H\big]  \nonumber\\
		&=1-\left(1-\frac{\eps}{n}\right)^n-n\cdot\frac\eps{n}\left(1-\frac\eps{n}\right)^{n-1}
\end{align*}
The inequality $(1-x)^n\leq 1-nx+{n\choose2}x^2$, holds for every $x>0$ and natural number $n$. Therefore,
\begin{align*}
		\Pr\big[\smax(X_1,\ldots,X_n)\geq H\big] \geq 1-\left(1-\eps+\frac{\eps^2}{2}\cdot\frac{n-1}{n}\right)-\eps\left(1-\eps\cdot\frac{n-1}{n}+\left(\frac{\eps}{n}\right)^2{n-1\choose2}\right)
		\geq \frac{\eps^2}{2}\cdot\frac{n-1}{n}
\end{align*}
By our assumption that $H>4\opt/\eps^3$, it follows that
$$
	\expct\big[\smax(X_1,\ldots,X_n)]\geq 
		H\cdot\Pr\big[\smax(X_1,\ldots,X_n)\geq H\big]\geq
		4\cdot\frac\opt{\eps^3}\cdot\left(\frac{\eps^2}{2}\cdot\frac{n-1}{n}\right)>\frac\opt\eps
$$
where the last inequality holds for any $n\geq2$. Thus, we obtain that $\sprice[(X_1,\ldots,X_n)]$ strictly exceeds $\opt$, for any $\eps\in(0,1)$, in contradiction to the definition of $\opt$.
\end{proof}

By Lemma~\ref{lem:UB_of_H} and Equation~\eqref{eq:defn_of_v_i} the number of rounded middle values is
\begin{equation}\label{eq:m<16/eps^4}
  	m\leq\frac{H}{\eps\opt}\leq\frac{4\frac{\opt}{\eps^3}}{\eps\opt}=\frac{4}{\eps^4}
\end{equation}
The next lemma shows that by rounding the middle values we lose a small fraction of $\midsprice[\Eopt]$. Let $\Eopt'$ denote the economy of rounded distributions $\D'_1,\ldots,\D'_n$.

\begin{lemma}\label{lem:mid_rounded_values_approx}
	
\begin{equation*}
  	\midsprice[\Eopt]\geq\midsprice[\Eopt']\geq\midsprice[\Eopt]-\eps\opt
\end{equation*}
\end{lemma}

\begin{proof}
Let $X_1,\ldots,X_n$ be values sampled from distributions $\D_1,\ldots,\D_n$, respectively. 
Map $X_i$ to $X'_i$, for each $i\in[n]$, via rounding, as follows:
$$
	X'_i=\begin{cases}
			0,& \quad\text{if $X_i<L$;}\\
			v_j\,(=L+j\cdot\eps\opt),& \quad\text{if $X_i\in[v_j,v_{j+1})$;}\\
			H,& \quad\text{if $X_i\geq H$.}
		\end{cases}
$$
Notice that $X'_1,\ldots,X'_n$ are distributed according to $\D'_1,\ldots,\D'_n$. Hence our mapping describes a correspondence between instances of $\Eopt$ and instances of $\Eopt'$. Clearly, by our mapping, whenever $X_i\in[L,H]$ it follows that
$X_i\geq X'_i \geq X_i - \eps\opt$. In particular, in each middle value instance of $\Eopt'$, the second-highest bidder value must satisfy the above inequality with the corresponding bidder in the corresponding instance of $\Eopt$. This implies the lemma.
\end{proof}

In $\Eopt'$, all distributions are discrete and support at most $m$ nonzero values, thus each distribution can be viewed as a vector of $m$ probabilities. After rounding the values we round these probabilities as well, to ensure that the number of such possible vectors is a function of $\eps$, independent of $n$.

Recall that we defined $p_{i,j}=\Pr_{V'_i\sim \D'_i}\big[V'_i=v_j]=\Pr_{V_i\sim\D_i}\big[v_j\leq V_i<v_{j+1}\big]$, for each $i\in[n]$ and $j\in[m]$. For $\gamma>0$, a constant to be determined later, define $t_{i,j}$ such that
$$
		p_{i,j}=\gamma\cdot t_{i,j}
$$ 
for each $i\in[n]$ and $j\in[m]$. Let $\hatt_{i,j}=\lfloor t_{i,j}\rfloor$ and define
$$
	\hatp_{i,j}=\gamma\cdot\hatt_{i,j}
$$
as the rounded down probability. Define $\D''_i$ as the distribution $\D'_i$ after rounding the probability of obtaining $v_j$ from $p_{i,j}$ down to $\hatp_{i,j}$, as described above, for each $i\in[n]$ and $j\in[m]$. The next lemma shows that by rounding the probabilities we lose only a small portion of the revenue. Let $\Eopt''$ denote the economy of rounded distributions $\D''_1,\ldots,\D''_n$.

\begin{lemma}\label{lem:rounding_probs_LB}
If $\gamma=\frac{\eps^8}{16n}$, then
$$\midsprice[\Eopt']+\frac{\eps^5}{4}\cdot\opt\geq	\midsprice[\Eopt'']\geq\, \midsprice[\Eopt']-\eps\cdot\opt$$
\end{lemma}
\begin{proof}
Recall that $v_j=L+ j\cdot\eps\opt$, for $j\in[m]$ as defined in Equation~\eqref{eq:defn_of_v_i}. Slightly abusing notation, denote $v_{m+1} = \infty$.
For every $i\in[n]$, let $F_i$ be the CDF of $\D_i$. We define the random variables $X'_i$ and $X''_i$, for each $i\in[n]$ as follows:
\begin{enumerate}
  \item Sample $q_i\in[0,1]$ uniformly at random.
  \item If $q_i<F_i(L)$, then set $X’_i = 0$ and $X''_i = 0$.
  \item Else,
  	\begin{itemize}
  		\item Let $j\in[m]$ satisfy $F_i(v_j)\leq q_i < F_i(v_{j+1})$.
  		\item Set $X'_i=v_j$.
  		\item If $F_i(v_j)\leq q_i < F_i(v_j)+\hatp_{i,j}$, set $X''_i=v_j$; otherwise $X''_i=0$. 
	\end{itemize}
\end{enumerate}

It is not hard to verify that $\Pr[X'_i=v_j]=F_i(v_{j+1})-F_i(v_j)=p_{i,j}$ and $\Pr[X''_i=v_j]=\hatp_{i,j}$, for every $j\in[m]$. Thus,
the variables $X'_1,\ldots,X'_n$ are distributed according to distributions  $\D'_1,\ldots,\D'_n$, respectively, and the variables $X''_1,\ldots,X''_n$ are distributed according to distributions $\D''_1,\ldots,\D''_n$, respectively. 
Hence the procedure defines a correspondence between instances of $\Eopt'$ and instances of $\Eopt''$. We may now compare the two economies according to their corresponding instances.

In every instance where $X''_i=X'_i$, for all $i$, the revenue of $\Eopt''$ and $\Eopt'$ is the same. We next show that this event occurs with very high probability.

Notice that $\hatp_{i,j}=\gamma\cdot\lfloor t_{i,j}\rfloor\geq \gamma\cdot(t_{i,j}-1)\geq p_{i,j}-\gamma$. Given that the support of $\D''_i$ is of size at most $m$, the total probability mass lost when rounding $\D'_i$ down to $\D''_i$ is at most $m\cdot\gamma$, for every $i$. Hence, $\Pr\big[X''_i\neq X'_i\big]\leq m\cdot\gamma$, for every $i\in[n]$. Therefore, by the union bound,
$$
	\Pr\big[\exists i: X''_i\neq X'_i\big]\leq \sum_{i=1}^{n}\Pr\big[ X''_i\neq X'_i\big]\leq n\cdot m \cdot \gamma
$$
Conservatively assume the revenue is zero if even one bidder has changed in the transition from $\Eopt'$ to $\Eopt''$. Since by definition $X'_i\leq H$, for every $i$, this assumption implies
\begin{align}
  	\midsprice[\Eopt'']\geq\, \midsprice[\Eopt']-\Pr\big[\exists i: X''_i\neq X'_i\big]\cdot H 
  		\geq\, \midsprice[\Eopt']-n\cdot m\cdot\gamma\cdot H \label{eq:lost_rev_from_rounding_probs}
\end{align}
We shall choose $\gamma$ such that $n\cdot m\cdot\gamma\cdot H\leq \eps\opt$. Plugging the values of $H$ and $m$ (see Lemma~\ref{lem:UB_of_H} and Equation~\eqref{eq:m<16/eps^4}) in this inequality we obtain
$$
	n\cdot\frac{4}{\eps^4}\cdot4\frac\opt{\eps^3}\cdot\gamma\leq \eps\opt
$$
Rearranging, we conclude that it suffices to set
$
	\gamma=\frac{\eps^8}{16n}
$
, as claimed.
Plugging this into Equation~\eqref{eq:lost_rev_from_rounding_probs}:
\begin{equation}\label{eq:rounding_probs_LB}
	\midsprice[\Eopt'']\geq\, \midsprice[\Eopt']-\eps\cdot\opt
\end{equation}
At first glance, it would appear that $\midsprice[\Eopt']\geq \midsprice[\Eopt'']$, since we only remove mass when rounding the probabilities. However, as we now argue, that is not necessarily true, and we prove the weaker inequality
$$
	\midsprice[\Eopt']\geq \midsprice[\Eopt'']-\frac{\eps^5}{4}\cdot\opt
$$
This issue stems from the fact that only instances with at most one high value bidder contribute to $\midsprice$, while instances with two or more high value bidders contribute to $\highsprice$ instead, for every economy $\econ$. Note that an instance with several high value bidders in $\Eopt'$ may be rounded down to an instance with no high value bidders in $\Eopt''$. The former instance contributes to $\highsprice[\Eopt']$ and has zero contribution to $\midsprice[\Eopt']$, whereas the latter instance has non-zero contribution to $\midsprice[\Eopt'']$. Due to such instances it is possible that $\midsprice[\Eopt']< \midsprice[\Eopt'']$. Therefore, it suffices to eliminate cases where \emph{any} high value bidder is rounded down to zero in $\Eopt''$, and the revenue from each instance that contributes to $\midsprice[\Eopt'']$ is at most $H$. This translates to the following inequality:
$$
		\midsprice[\Eopt']\geq\, \midsprice[\Eopt'']- \Pr\big[\exists i : (X'_i=H)\wedge (X_i''=0)\big]\cdot H
$$ 
For each $i$, the probability that $X_i'=H$ and $X_i''=0$ is at most $\gamma$, and thus, by the union bound, summing over all bidders we obtain a probability of $n\cdot\gamma$. Hence, 
\begin{align}
	\midsprice[\Eopt']\geq\, 	\midsprice[\Eopt'']- n\cdot\gamma\cdot H \nonumber
					\geq\, \midsprice[\Eopt'']-\frac{\eps^5}{4}\cdot\opt \nonumber
\end{align}
where the second inequality follows by 
Lemma~\ref{lem:UB_of_H}. This proves the lemma.
\end{proof}

Combining Lemmata~\ref{lem:mid_rounded_values_approx} and~\ref{lem:rounding_probs_LB}, gives us
$$
	\midsprice[\Eopt]+\frac{\eps^5}{4}\cdot\opt\geq \midsprice[\Eopt'']\geq \midsprice[\Eopt]-2\eps\cdot\opt
$$
We obtain the following lemma as an immediate conclusion.
\begin{lemma}\label{lem:mid_value_approx}
	Let $\econ$ be an economy where the rounding described above (of values and probabilities) yields $\econ''=\Eopt''$, i.e., the values $\hatp_{i,j}$ are identical in the two economies, for all $i\in[n],j\in[m]$. Then,
	$$
		\midsprice\geq \midsprice[\Eopt]-\left(2\eps+\frac{\eps^5}{4}\right)\opt
	$$	
\end{lemma}

Finally, we separate the set of distributions in $\Eopt$ into groups
$G_1,\ldots,G_k$ according to their rounded representation. That is, a pair of distributions $\D_i$ and $\D_j$, for $i,j\in[n]$, belong to the same group if $\D''_i\equiv\D''_j$. 

The number of groups is upper bounded by the number of possible rounded probability vectors. The following lemma shows that this bound depends only on $\eps$.

\begin{lemma}\label{lem:k<Exp(1/eps)}
	The number of groups, denoted by $k$, satisfies $k\leq\left(\frac{4}{\eps^4}\right)^{\frac{16}{\eps^{10}}}$.
\end{lemma}
\begin{proof}
	Each group is represented by a distinct probability vector. For each $i\in[n]$, the rounded probability vector of distribution $\D_i$ is $(\hatp_{i,j})_{j=1}^m$. It contains $m$ entries and each entry is upper bounded as follows:
	$$\hatp_{i,j}\leq p_{i,j}=\Pr_{V_i\sim  \D_i}\big[v_j\leq V_i<v_{j+1}\big]\leq \Pr_{V_i\sim\D_i}\big[V_i\geq L\big]\leq \frac{1}{\eps^2n} $$
where the last inequality is due to the definition of $L$. Additionally, $\hatp_{i,j}=\gamma\cdot\hatt_{i,j}$, so we have
$$
	\hatt_{i,j}\leq\frac{1}{\gamma\cdot\eps^2n}\leq \frac{16}{\eps^{10}}
$$
where the value of $\gamma$ is from Lemma \ref{lem:rounding_probs_LB}.
Given that $\hatt_{i,j}$ is a non-negative integer for every $i$ and $j$, it follows that the number of possible vectors of the form $(\gamma\cdot\hatt_{i,j})_{j=1}^m$ is at most $m^{\frac{16}{\eps^{10}}}$.

By the upper bound on $m$ of Equation~\eqref{eq:m<16/eps^4}, and since each group is represented by a distinct probability vector, we conclude that $k$, the number of groups, is at most
$
	k\leq \left(\frac{4}{\eps^4}\right)^{\frac{16}{\eps^{10}}}
$
as claimed.
\end{proof}

When constructing $\Eeps$, we keep each bidder in the group as in $\Eopt$, i.e., if bidder $i$ is drawn from $\D_i\in G$ in $\Eopt$, for some group $G$, we draw bidder $i$ from $\D'_i$ in $\Eeps$ such that $\D'_i$ belongs to the same group $G$. Thus, Lemma~\ref{lem:mid_value_approx} ensures we approximate the revenue of the middle values.

\subsection{Approximating the High Values}

If group assignments are maintained, the contribution of middle values in $\Eopt$ is well approximated. Moreover, in any instance where $\SP>H$, the probability mass of non-high values has no influence on the revenue. Then, when focusing on high value instances, we may truncate all distributions, setting the probability mass below $H$ to zero. Let $\tD_1,\ldots,\tD_n$ be the ``truncated'' distributions.

We now consider the contribution of every possible (unordered) pair of distributions, by restricting our focus to instances where a specific pair of bidders $i$ and $j$ have drawn the two largest values.
Let $B_{i,j}$ denote the event where the values of bidders $i$ and $j$ are largest, for each $1\leq i<j\leq n$.
Define,
$$
	r_{i,j}=\expct_{V_i\sim\tD_i}\big[\smax(V_1,\ldots,V_n)\cdot\one_{B_{i,j}}\big]
$$
for $1\leq i<j\leq n$.
Note that each high value instance belongs to at least one $B_{i,j}$ event. Thus, 
\begin{equation}\label{eq:sp<sum_r_ij}
  \highsprice[\Eopt]=\expct_{V_i\sim\tD_i}\big[\smax(V_1,\ldots,V_n)\big] \leq \sum_{i<j}{r_{i,j}}
\end{equation}
Moreover, notice that for any given sample $V_1,\ldots,V_n$ from the distributions $\tD_1,\ldots,\tD_n$, the event $B_{i,j}$ implies that  the random variable $\smax(V_1,\ldots,V_n)$ is equal to the random variable $\smax(V_i,V_j) $. Thus,
\begin{equation}\label{eq:r_i,j<smax(D_i,D_j)(v2)}
  r_{i,j}=\expct_{V_\ell\sim\tD_\ell}\big[\smax(V_1,\ldots,V_n)\cdot\one_{B_{i,j}}\big]
  		 =\expct_{V_\ell\sim\tD_\ell}\big[\smax(V_i,V_j)\cdot\one_{B_{i,j}}\big]
  		 \leq \expct_{V_\ell\sim\tD_\ell}\big[\smax(V_i,V_j)\big]
\end{equation}

Let the set of groups of bidder value distribution determined by $\Eopt$, as described in the previous subsection, be $G_1,\ldots,G_k$.

We call a group $G$ \emph{big} if $|G|>\frac{1}{\eps}$, otherwise we refer to it as \emph{small}. Slightly abusing notation, we say that a bidder $i$ is in group $G$ if the distribution $\D_i$ (from which $i$ draws her value) is in $G$.  We construct an economy $\Eeps'$ by the following algorithm:

\begin{center}
\begin{tcolorbox}[colback=white ,
                  colframe=black,
                  width=0.9\linewidth,
                 ]
  	\begin{enumerate}
 	 	\item Keep all bidders from \emph{small} groups the same as in $\Eopt$.
 	 	\item For each \emph{big} group $G$, let $T=\frac{1}{\eps}$.
 	 	\begin{enumerate}
  			\item For each bidder $i$ in $G$, draw an index $s_i\in[T]$ uniformly at random.
  			\item Sample $T$ distributions from $G$ uniformly at random (allowing repetition). Denote the sampled distributions $\DG_1,\ldots,\DG_T. $
			\item Assign each bidder $i \in [n]$ the distribution $\DG_{s_i}$.
 
		\end{enumerate}

	\end{enumerate}


\end{tcolorbox}
\end{center}

\noindent\textbf{Number of distributions used:} First note that this algorithm uses at most $\frac{1}{\eps}$ distributions from each group, and there are $k\leq\left(\frac{4}{\eps^4}\right)^{\frac{16}{\eps^{10}}}$ groups. So the number of distributions used in total is a function of $\eps$ only, as desired.

\noindent\textbf{Proof of correctness:} The rest of this section is dedicated to showing that the approximation factor achieved by the economy $\Eeps'$ (as constructed by the algorithm) is in fact $1-\eps$.

Let $X_1,\ldots,X_n$ be the sampled values of the bidders in economy $\Eeps'$. Let $Z_{i,j}$, for $1\leq i<j\leq n$, denote the event where all bidders other than bidders $i$ and $j$ have zero value in $\Eeps'$. That is, $X_a=0$, for all $a\in[n]\setminus\{i,j\}$. Recall that we discarded all the probability mass below $H$, so by the definition of $H$ the probability that any bidder value is zero is at least $1-\frac\eps{n}$. Hence, for every $1\leq i<j\leq n$,
\begin{equation}\label{eq:Z_ij(epsilon)}
  \Pr\big[Z_{i,j}\big]\geq \left(1-\frac\eps{n}\right)^{n-2}
  				\geq \;\left(1-\frac\eps{n}\right)^{n}  
  				\;\geq\; 1-\eps
\end{equation}
We thus obtain,
\begin{align}
	\highsprice[\Eeps'] =&\;\; \expct\big[\smax(X_1,\ldots,X_n)\big] &\nonumber\\
	\geq&\;\;\expct\bigg[\smax(X_1,\ldots,X_n)\cdot\one\bigg\{\bigvee_{i<j}Z_{i,j}\bigg\}\bigg]  &\nonumber \\
	 \;=&\;\; \sum_{i<j}\expct\left[\smax(X_1,\ldots,X_n)\cdot\one_{Z_{i,j}}\right]  &\text{($Z_{i,j}$ are disjoint events)}\nonumber \\
	=&\;\; \sum_{i<j}\expct\left[\smax(X_i,X_j)\cdot\one_{Z_{i,j}}\right]  &\text{(by definition of $Z_{i,j}$)}\   \nonumber\\
	=&\;\; \sum_{i<j}\expct\left[\smax(X_i,X_j)\right]\cdot\Pr[Z_{i,j}]   &\text{(independent r.v.'s)}\nonumber\\
	\geq&\;\; (1-\eps)\sum_{i<j}\expct\left[\smax(X_i,X_j)\right] &\text{(by Eq.
	 ~\eqref{eq:Z_ij(epsilon)})}  \label{eq:high_eps_LB} 
\end{align}

Let $\algdist$ denote a construction generated by the algorithm ($\algdist$ is a random variable, over which we will later take expectation). 
We consider the number of times each pair of distributions have been assigned to bidders by our algorithm.
Formally, define the following random variable, for each $a,b\in[n]$,
$$
	n_{a,b}= \lvert\{(X_i,X_j)\mid \text{$i,j\in[n]$: $X_i$ and $X_j$ were drawn from $\tD_a$ and $\tD_b$, , respectively, given $\algdist$} \}\rvert	
$$
Given that the construction $\algdist$ was generated by the algorithm, Equation~\eqref{eq:high_eps_LB} can now be written as
\begin{align*}
	\highsprice[\Eeps'] 	\geq&\;\; (1-\eps)\sum_{i<j}\expct\left[\smax(X_i,X_j)\right]~  &\nonumber \\
			=&\;\; (1-\eps)\sum_{a<b}n_{a,b}\cdot\expct_{V_\ell\sim\tD_\ell}\left[\smax(V_a,V_b)\right] &\text{(rearranging sum by distributions)}  \nonumber \\
	 \;=&\;\;  (1-\eps)\sum_{a<b}n_{a,b}\cdot r_{a,b} &\text{(by Equation~\eqref{eq:r_i,j<smax(D_i,D_j)(v2)})}\nonumber 
\end{align*}

Taking expectation over $\algdist$, by linearity of expectation, we obtain
\begin{equation}\label{eq:E_S<(1-eps)sum_E_S[n_a,b]*r_a,b}
\expct_{\algdist}\big[\highsprice[\Eeps']\big]\geq (1-\eps)\sum_{a<b}\expct_{\algdist}\big[n_{a,b}\big]\cdot r_{a,b}
\end{equation}

The next lemma shows that on the right hand side of the above equation the sum is very close to the sum $\sum_{i<j}{r_{i,j}}$, which is an upper bound of $\highsprice[\Eopt]$, by Equation~\eqref{eq:sp<sum_r_ij}.
\begin{lemma}\label{lem:E[n_a,b]>1-eps}
	For every $1\leq a < b \leq n$,
	$$
		\expct_{\algdist}\big[n_{a,b}\big]\geq (1-2\eps)\,.
	$$
\end{lemma}
\begin{proof}
Let $N_a$ (resp., $N_b$) denote the number of bidders assigned distribution $\tD_a$ (resp., $\tD_b$) in the construction $\algdist$. Recall that $n_{a,b}$ denotes the number of ordered pairs of bidders $(i,j)$ such that  $(X_i,X_j)\sim(\tD_a,\tD_b)$, in the construction $\algdist$. Therefore, $n_{a,b}=N_a\cdot N_b$. We start by proving the following claim.

\begin{claim}\label{clm:E[N_a]=1}
	$\expct_{\algdist}\big[N_a\big]=\expct_{\algdist}\big[N_b\big]=1.$
\end{claim}
\begin{proof}
	We prove the claim for $N_a$, the proof for $N_b$ is analogous. Let $G$ be the group that distribution $\tD_a$ belongs to. If $G$ is a small group, then bidder $a$ keeps the distribution $\tD_a$, regardless of $\algdist$, so $N_a=1$, and thus the claim holds. Hence, we may assume $G$ is a big group.  
	
	Consider a specific bidder $i$ in $G$. Let $B_i$ denote the event that $X_i\sim \tD_a$ in $\algdist$. For this event to occur, the sampled distribution $\DG_{s_i}$, where $s_i\in[T]$ is the index sampled for bidder $i$, must be equal to $\tD_a$. Regardless of the value $s_i$, the distribution $\DG_{s_i}$ is simply an element of the group $G$ sampled uniformly at random. Therefore, the probability that the sampled distribution is indeed $\tD_a$ is equal to $\frac1{|G|}$. It follows that $\Pr[B_i]=\frac{1}{|G|}$. Hence, by linearity of expectation 
	$$
	\expct_{\algdist}\big[N_a\big]=\sum_{i\in G}\expct_{\algdist}\big[\one_{B_i}\big]=\sum_{i\in G}\Pr_{\algdist}\big[B_i]=|G|\cdot\frac1{|G|}=1
	$$
\end{proof}

We are now ready to calculate $n_{a,b}$. Consider the case where $\tD_a$ and $\tD_b$ belong to different groups. Since the algorithm samples from each group separately, $N_a$ and $N_b$ are independent random variables. 
Thus, by Claim~\ref{clm:E[N_a]=1} and the fact that $n_{a,b}=N_a\cdot N_b$, we obtain
$$\expct_\algdist[n_{a,b}]=\expct_\algdist[N_a\cdot N_b]=\expct_{\algdist}[N_a]\cdot\expct_{\algdist}[N_b]=1$$
Thus, the lemma holds in this case.

Next, consider the case where $\tD_a$ and $\tD_b$ belong to the same group $G$, but the group $G$ is small. The algorithm assigns each of these distributions to exactly one bidder, and thus $n_{a,b}=1$. Hence, the lemma holds in this case as well.

It remains to consider the case where $\tD_a$ and $\tD_b$ belong to the same big group $G$. For every pair of bidders $i,j\in G$, we define the event
$$
	A_{i,j}=\{\text{the distributions chosen for bidders $i$ and $j$ are, respectively, $\tD_a$ and $\tD_b$}\}
$$
Consider the probability of this event. Consider the indices $s_i,s_j\in[T]$ sampled in construction $\algdist$. If $s_i=s_j$, which occurs with probability $\frac1T$, then it is not possible to have both $\DG_{s_i}=\tD_a$ and $\DG_{s_j}=\tD_b$, since $\DG_{s_i}$ and $\DG_{s_j}$ are equivalent while $\tD_a$ and $\tD_b$ are two distinct distributions. Else, when $s_i\neq s_j$, the samples $\DG_{s_i}$ and $\DG_{s_j}$ are two independent samples from $G$ and thus the probability that $(\DG_{s_i},\DG_{s_j}) = (\tD_a,\tD_b)$ is exactly $\frac1{|G|}\cdot\frac1{|G|}$. We conclude from this discussion that
$$
	\Pr\big[A_{i,j}\big]=\left(1-\frac1T\right)\cdot\frac1{|G|^2}=(1-\eps)\cdot\frac1{|G|^2}
$$
where the second equality is by the definition of $T$.

Summing over all (ordered) pairs of bidders $i$,$j$ in group $G$ we obtain
$$
	n_{a,b}=\sum_{i\neq j}\one_{A_{i,j}}
$$
Taking expectation over $\algdist$, by linearity of expectation, we obtain
$$
	\expct_\algdist[n_{a,b}]=\sum_{i\neq j}\Pr\big[A_{i,j}\big]
	= (1-\eps)\cdot\sum_{i\neq j}\frac1{|G|^2}= (1-\eps)\frac1{|G|^2}\cdot|G|\cdot(|G|-1)
	=(1-\eps)\left(1-\frac1{|G|}\right)
$$
As $G$ is a big group, we have $\big(1-\frac1{|G|}\big)>1-\eps$. We thus obtain from the above, $\expct_\algdist[n_{a,b}]\geq (1-\eps)^2\geq(1-2\eps)$. This proves the lemma.
\end{proof}

It now follows that $\Eeps'$ generates almost all of the contribution of high value instances in $\Eopt$. This is shown by plugging Lemma~\ref{lem:E[n_a,b]>1-eps} into Equation~\eqref{eq:E_S<(1-eps)sum_E_S[n_a,b]*r_a,b}, which yields
$$
	\expct_\algdist\big[\highsprice[\Eeps']\big]\geq 
	(1-\eps)\sum_{a<b} \expct_{\algdist}\big[n_{a,b}\big] r_{a,b}
	\geq (1-\eps)(1-2\eps)\sum_{a<b} r_{a,b}
	\geq (1-3\eps)\cdot\highsprice[\Eopt]
$$
where the last inequality follows from Equation~\eqref{eq:sp<sum_r_ij}.

Finally, since the inequality above holds in expectation over the set $\algdist$, there must exist a specific set $\algdist$ for which
$$
\highsprice[\Eeps']\geq 
	 (1-3\eps)\cdot\highsprice[\Eopt]
$$
\subsection{Putting it all Together}
%
By construction, for every $i\in [n]$, bidder $i$ in $\Eeps'$ belongs to the same group to which bidder $i$ in $\Eopt$ belongs. Recalling that distributions within the same group have the same ``discretized'' rounded representation of middle values, Lemma~\ref{lem:mid_value_approx} now implies
%
$$
	\midsprice[\Eeps']\geq \midsprice[\Eopt]-\left(2\eps+\frac{\eps^5}{4}\right)\opt
$$
To cover the low values, we add $\eps\cdot n$ bidders to $\Eeps'$ with values drawn from $\D_L$, obtaining a larger economy $\Eeps''$ containing $(1+\eps)n$ bidders. By Lemma~\ref{lem:low_value_approx},
$$
	\lowsprice[\Eeps'']\geq (1-3\eps)\lowsprice[\Eopt]
$$
Recalling that $\highsprice[\Eeps']\geq 
	 (1-3\eps)\cdot\highsprice[\Eopt]$, we obtain:
\begin{align*}
	\sprice[\Eeps'']\;&\geq\; (1-3\eps)\lowsprice[\Eopt]
						+ \midsprice[\Eopt]-\left(2\eps+\frac{\eps^5}{4}\right)\opt
						+ (1-3\eps)\highsprice[\Eopt]\\
					&\geq\; (1-3\eps)\sprice[\Eopt]-	\left(2\eps+\frac{\eps^5}{4}\right)\\
					&\geq\; (1-6\eps)\opt
\end{align*}
where the last inequality is due to $\sprice[\Eopt]=\opt$. Finally, it remains to ``shrink'' the economy $\Eeps''$ back to an $n$ bidder economy. By Lemma~\ref{lem:close_to_OPT_without_eps*n_bidders} there exists an economy $\Eeps$ consisting of $n$ bidders from $\Eeps''$ such that
$$
	\sprice[\Eeps]\geq (1-2\eps)\sprice[\Eeps']\geq (1-2\eps)(1-6\eps)\opt \geq (1-8\eps)\opt
$$
This shows that the constructed economy $\Eeps$ attains a $(1-8\eps)$-fraction of the ideal second price revenue, as required. Moreover, in the construction of $\Eeps$ we used at most $\frac1{\eps}$ distributions per group, and the number of groups is bounded by Lemma~\ref{lem:k<Exp(1/eps)}, so the number of distributions used is  at most
$$
	\frac{1}{\eps}\cdot\left(\frac{4}{\eps^4}\right)^{\frac{16}{\eps^{10}}}=\left(\frac1\eps\right)^{O\left(\frac1{\eps^{10}}\right)}
$$
This concludes the proof of the theorem.

	\section*{Acknowledgments}
		The first and third authors were supported by BSF grant 2016192 and ISF grant 2185/19. The second author was supported by the European Research Council (ERC) under the European Union's Horizon 2020 research and innovation program (grant agreement No. 866132), and by the Israel Science Foundation (grant number 317/17).
\else

	\noindent\rule{8cm}{0.4pt}

	\noindent{\footnotesize
		The first and third authors were supported by BSF grant 2016192 and ISF grant 2185/19. The second author was supported by the European Research Council (ERC) under the European Union's Horizon 2020 research and innovation program (grant agreement No. 866132), and by the Israel Science Foundation (grant number 317/17).}
\fi

\bibliographystyle{plain}
\bibliography{simple}

\iffull
	\appendix
	\section{An Impossibility Result for Regular Distributions}\label{app-regular}

As already mentioned in the introduction, there exists a set of regular distributions $D=\{\mathcal D_1,\ldots, \mathcal D_n\}$ such that if $\mathbf E$ is a homogeneous economy where the value of each bidder is drawn from some $\mathcal D\in D$, then the optimal auction for $\mathbf E$ generates about half of the ideal revenue. We now slightly strengthen this result and show that a similar bound holds even if all distributions in $D$ are regular. Recall that a distribution with CDF $F$ and PDF $f$ that is defined on $[a,b]$ is regular if its virtual value $\Phi(v)=v-\frac {1-F(v)} {f(v)}$ is non-decreasing in $[a,b]$.

\begin{theorem}
For every $\eps>0$ and $n>2$, there exists a set of regular distributions $ D=\{\mathcal D_1,\ldots, \mathcal D_n\}$ such that if  $\,\mathbf E$ is a homogeneous economy, where the value of each bidder is drawn from some $\mathcal D\in D$, then the optimal auction for $\mathbf E$ generates at most a $\frac {n} {2n-1}+\eps$ fraction of the ideal revenue.
\end{theorem}

Recall that a distribution with CDF $F$ and PDF $f$ is called \emph{equal revenue} in the range $[1, h]$ if for each $1<x< h$ it holds that $F(x)=1-\frac 1 x$ and $f(h)=\frac 1 h$. The basic property of an equal revenue distribution is that the maximal revenue that can be extracted from a bidder whose value is drawn from that distribution is $1$: the revenue of any take-it-or-leave-it offer $p$, $1\leq p\leq h$, is $1$ (the bidder accepts and pays $p$ with probability $\frac 1 p$). The revenue of any other offer is strictly smaller. One can verify that the virtual value of an equal revenue distribution is constant, hence it is regular.

Let $D$ contain two distributions: the equal revenue distribution $\mathcal D_{ER}$ on $[1,h]$, for some $h \gg n$, and the constant distribution $\mathcal D_c$ that returns a value $n$ with probability $1$. Note that both distributions are regular.

To give a lower bound on the ideal revenue of $D$, consider the economy $\mathbf E$ that contains one bidder with distribution $\mathcal D_c$ and $n-1$ bidders with distributions $\mathcal D_{ER}$. Consider the following (obviously non-optimal) auction for $\mathbf E$: if there is exactly one bidder with value $h$, then this bidder gets the item and pays $h$. If there are two or more bidders with value $h$, none of the bidders gets the item. Else, the bidder with the constant distribution gets the item and pays $n$. We now analyze the revenue of this auction.

The probability that none of the bidders has value $h$ is, by the union bound, at least $1-\frac {n-1} h$. In this case the bidder with the constant distribution gets the item and pays $n$.

Each of the other bidders essentially faces a take-it-or-leave-it auction with price $h$, unless one of the other bidders has value $h$, in which case the item is not allocated. Since we consider equal revenue distributions and the 
bidder values are independent, using the union bound again the expected revenue is $(n-1) \cdot \Pr[\text{none of the other bidders has value $h$}]\geq (n-1)\cdot (1-\frac {n-2} h)$.

Thus, by linearity of expectation, the total revenue of the auction is at least $n\cdot (1- \frac {n-1} h) + (n-1)\cdot (1-\frac {n-2} h)$ which approaches $2n-1$ as $h$ increases.

Finally, to complete the proof we observe that the optimal revenue of any homogeneous economy is at most $n$: if the distribution of all bidders is drawn from $\mathcal D_C$ then the optimal revenue is obviously $n$. If all distributions are equal revenue then the expected payment of each bidder is at most $1$, thus the optimal revenue is bounded by $n$ in this case as well.
	\section{Missing Proofs}\label{apx:missing-proofs}
		\CLM22proof
\fi

\end{document}